\documentclass{article}
\usepackage{amsmath,amssymb}
\usepackage{authblk}
\usepackage{cite}
\usepackage{comment}

\makeatletter
\@addtoreset{equation}{section}

\makeatother

\newcommand{\D}{\mathcal D}
\newcommand{\hc}{\mathrm{h.c.}}
\newcommand{\del}{\partial}

\newcommand{\Tr}{\mathrm{Tr}}

\newcommand{\E}{\mathcal E}
\newcommand{\W}{\mathcal W}

\title{\textbf{Anomaly Calculation by Path Integral in Superspace}}
\author{Akihisa D.-E. Tateishi}
\affil{\it Department of Physics, The University of Tokyo, Tokyo 113-0033, Japan}
\date{}

\begin{document}
\maketitle

\begin{abstract}
A direct anomaly calculation method for general supersymmetric models by path integral measure in superspace is established. It includes the traditional Konishi anomaly as a specific case. As another example, superconformal anomaly in $\mathcal N =1, D=4$ conformal supergravity is calculated.
\end{abstract}

\newpage

\section{Introduction}
Anomaly calculation is one of the most instructive topics to comprehend quantum theories. For example, a brilliant way of anomaly calculation utilising path integral measure was proposed by Fujikawa\cite{1}. However, anomalies in supersymmetric theories are in general hard to calculate due to the complexity of its Lagrangian, and sometimes are not derived in a manifestly supersymmetric way. In this paper, we first review a supersymmetric extention\cite{2} of Fujikawa method and sophisticate its vague point to check that it reproduces the traditional Konishi anomaly, the chiral anomaly in $\mathcal N=1$ supersymmetric gauge theory, which was originally derived in a supergraphical way\cite{3}. As another example, we also calculate superconformal anomaly in $\mathcal N=1, D=4$ conformal supergravity. In previous researches \cite{01,02} they calculate the superconformal anomaly in Poincar\'e supergravity, and this paper is the conformal SUGRA version of it.\par
This paper is organised as follows. In section 2 we review Fujikawa method and extend it to superspace. In section 3 we review $\mathcal N=1, D=4$ conformal supergravity described in conformal superspace formalism. In section 4, the main part of this paper, we calculate superconformal anomaly in $\mathcal N=1, D=4$ conformal supergravity. Section 5 is the summary.

\section{Fujikawa Method in Superspace\label{section_Fujikawa_method}}
In this section we briefly review a powerful anomaly-calculation method by Fujikawa\cite{1} and its supersymmetric extension by Konishi and Shizuya\cite{2}. Thereafter we sophisticate and generalise the Konishi-Shizuya method to a more useful form.\par
Fujikawa method is a technique to calculate anomaly by variation of path integral measure. In general 4-dimensional Euclidean theories, consider infinitesimal transformation of a field variable $\phi$ in path integral:
\begin{align}
Z=&\int\D\phi\,e^{-S[\phi]}\\
=&\int\D\phi'\,e^{-S[\phi']}\\
=&\int\D\phi\,e^Je^{-S[\phi]-\Delta S},\label{infinitesimal_field_redifinition}
\end{align}
where $\phi'=\phi+A\phi$ and $A$ is an arbitrary infinitesimal (in general operatorial) transformation parameter, $e^J$ the Jacobian for the field redifinition, symbolically $e^J=\det(\del\phi'/\del\phi)$, and $\Delta S$ the variation of the action, which contains the conserved quantity $Q[\phi]$ as
\begin{align}
\Delta S=\int d^4x\,A\,Q[\phi].\label{definition_conserved_quantity}
\end{align}
The explicit form of the Jacobian $e^J$ is given by functional trace\cite{1,4}:
\begin{align}
J=\int d^4x\,A\,T,\label{jacobian_by_trace}
\end{align}
where $T$ is the diagonal entry regularised by a `covariant differential operator' $\mathcal O$, which respects a more significant symmetry than the anomalous one:
\begin{align}
T&=\left.\exp\left(\frac{\mathcal O_x^2}{M^2}\right)\delta^4(x-y)\right|_{y\to x,M\to\infty}\\
&=\left.\int\frac{d^4k}{(2\pi)^4}\,e^{-ikx}\exp\left(\frac{\mathcal O^2}{M^2}\right)e^{ikx}\right|_{M\to\infty}\label{non-susy_fujikawa}.
\end{align}
From (\ref{infinitesimal_field_redifinition}),(\ref{definition_conserved_quantity}), and (\ref{jacobian_by_trace}), the expectation value of the classical conserved quantity is finally expressed by the functional trace:
\begin{align}
\langle Q\rangle=T,\label{anomaly_formula_general}
\end{align}
since the transformation parameter $A$ is arbitrary.\par
Next, we extend this formulation to superspace\footnote{Superspace notation follows\cite{5,6} throughout this paper.}. Here we specialise in the case where the field variable of path integral is a chiral superfield $\Phi$. Then (\ref{definition_conserved_quantity}) and (\ref{anomaly_formula_general}) are straightforwardly extended to superspace:
\begin{align}
\Delta S=\int d^8z\,A\,Q[\Phi],&\quad \Phi\mapsto\Phi'=\Phi+A\,\Phi,\\
\langle Q\rangle&=T,
\end{align}
and the functional trace $T$ should be understood the functional trace in the chiral subspace:
\begin{align}
T&=\left.\exp\left(\frac{\mathcal O_z^2}{M^2}\right)\delta^6(z_L-z'_L)\right|_{z'\to z,M\to\infty}\\
&=\left.\int\frac{d^6w}{4\pi^4}e^{-iwz_L}\exp\left(\frac{\mathcal O^2}{M^2}\right)e^{iwz_L}\right|_{M\to\infty},\label{supertrace_formula}
\end{align}
where $z_L=(x_L,\theta)\equiv(x+i\theta\sigma\bar\theta,\theta)$, $w=(k,\varphi)$, $d^6w=d^4kd^2\varphi$ with $\varphi$ fermionic, and we have used\footnote{In the previous research \cite{2} they use a more complicated way to evaluate chiral functional trace, but this way is simpler, we believe.}
\begin{align}
\delta^2(\theta-\theta')=(\theta-\theta')^2=4\int d^2\varphi\,e^{i\varphi(\theta-\theta')}.
\end{align}
\par
Now we can check that the general formula (\ref{supertrace_formula}) reproduces the Konishi anomaly\cite{3}. The action for a chiral matter superfield $\Phi$,
\begin{align}
S=\int d^8z\,\bar\Phi\,e^V\Phi,
\end{align}
with super-Yang-Mills background $V$, is invariant under the gauge transformation:
\begin{align}
\Phi\mapsto e^\Sigma\Phi,\quad\bar\Phi\mapsto\bar\Phi\,e^{\bar\Sigma},\quad e^V\mapsto e^{-\bar\Sigma}e^Ve^{-\Sigma},
\end{align}
with a chiral (in general algebra-valued) parameter $\Sigma$. The action is also invariant under chiral rotation:
\begin{align}
\Phi\mapsto e^{i\lambda}\Phi,\quad \bar\Phi\mapsto e^{-i\lambda}\bar\Phi,
\end{align}
with a real constant parameter  $\lambda$. Promoting $i\lambda$ to a local chiral parameter $\Lambda$, we obtain the variation of the action:
\begin{align}
\Delta S=\int d^8z\,\Lambda\,\bar\Phi\,e^V\Phi+\hc
\end{align}
Thus the classical conserved quantity is
\begin{align}
Q=\frac{\Delta S}{\Delta \Lambda}=-\frac14\bar D^2(\bar\Phi\,e^V\Phi),
\end{align}
where we have used
\begin{align}
\frac\Delta{\Delta\Lambda}\int d^8z\,\Lambda\Psi=\frac\Delta{\Delta\Lambda}\int d^6z\,\Lambda\frac{-1}4\bar D^2\Psi=-\frac14\bar D^2\Psi,
\end{align}
for a chiral parameter $\Lambda$.
On the other hand, the gauge covariant operator $\mathcal O$ should be taken
\begin{align}
\mathcal O^2\Phi&=\frac1{16}\bar D^2e^{-V}D^2e^V\Phi\\
&=\left(\Box-\frac12W^\alpha D_\alpha+iY^a\del_a+X\right)\Phi,
\end{align}
since $\bar D_{\dot\alpha}\Phi=0$, where
\begin{align}
X\equiv\frac1{16}\bar D^2(e^{-V}D^2e^V),\quad Y^a\equiv-\frac12\sigma^a_{\alpha\dot\alpha}\bar D^{\dot\alpha}(e^{-V}D^\alpha e^V),
\end{align}
and we have used the Yang-Mills field strength
\begin{align}
W_\alpha\equiv-\frac14\bar D^2e^{-V}D_\alpha e^V.
\end{align}
Then the chiral functional trace (\ref{supertrace_formula}) is
\begin{align}
&\int\frac{d^6w}{4\pi^4}e^{-iwz_L}\exp\frac1{M^2}\left(\Box-\frac12W^\alpha D_\alpha+iY^a\del_a+X\right)e^{iwz_L}\\
=&\int\frac{d^6w}{4\pi^4}\exp\frac1{M^2}\bigg[(\del+ik)^2-\frac12W^\alpha \left(D+i\varphi-2k_a\sigma^a\bar\theta\right)_\alpha\nonumber\\
&\qquad\qquad\qquad\qquad+iY^a(\del+ik)_a+X\bigg]\label{trace_calculation_1}\\
=&\int\frac{d^6w}{4\pi^4}M^3\exp\bigg[\left(\frac\del M+ik\right)^2-\frac12W^\alpha \left(\frac D{M^2}+i\frac\varphi{M^{3/2}}-2\frac{k_a}M\sigma^a\bar\theta\right)_\alpha\nonumber\\
&\qquad\qquad\qquad\qquad+iY^a\left(\frac\del{M^2}+i\frac kM\right)_a+\frac X{M^2}\bigg]\label{trace_calculation_2}\\
\xrightarrow{M\to\infty}&\int\frac{d^6w}{4\pi^4}e^{-k^2}\frac12\left(-\frac i2W^\alpha\varphi_\alpha\right)^2\label{trace_calculation_3}\\
=&\frac1{64\pi^2}W^\alpha W_\alpha,
\end{align}
where we have rescaled $(k,\varphi)\mapsto(Mk,\sqrt M\varphi)$ from (\ref{trace_calculation_1}) to (\ref{trace_calculation_2}), and used the fact from (\ref{trace_calculation_2}) to (\ref{trace_calculation_3}) that only the term with the factor $\varphi^2$ and a nonnegative power of $M$ survives the chiral integral $\int d^2\varphi$ and the cutoff limit $M\to\infty$. Finally the conservation law $\langle Q\rangle=T$ in the quantum theory is specifically rewritten\footnote{Note that implicit trace should be regarded at the end of the evaluation of (\ref{supertrace_formula}) if the theory has nonabelian gauge fields.}
\begin{align}
\frac{-1}4\bar D^2\left\langle\bar\Phi\,e^V\Phi\right\rangle=\frac1{64\pi^2}\Tr W^\alpha W_\alpha.\label{Konishi_anomaly}
\end{align}
This surely is the Konishi anomaly\footnote{The prefactor $1/64\pi^2$ is sometimes replaced by $1/32\pi^2$ or $1/16\pi^2$ in some other papers. This is a mere conventional problem. See Appendix \ref{appendix_coefficient} for a detail.}, what we would like to derive.\par
Here are two advantages of this manifestly supersymmetric version of Fujikawa method. One is merely that the calculation becomes much easier: we only have to calculate the coefficient $W^\alpha$ out of $\mathcal O^2=\Box-(1/2)W^\alpha D_\alpha+\cdots$. The other is that the bosonic and fermionic divergences automatically cancels out. As for the non-SUSY functional trace (\ref{non-susy_fujikawa}), the cutoff limit $M\to\infty$ sometimes brings about divergences such as $M^2$ or $M^4$ from both bosonic and fermionic integrals, and thereafter they cancel each other out if the theory has supersymmetry. On the other hand the supersymmetric cancellation of divergences is automatic and obvious in (\ref{trace_calculation_3}) by superspace integration.

\section{$\mathcal N=1, D=4$ Conformal Supergravity}
In this section we review $\mathcal N=1, D=4$ conformal supergravity. The results are quoted from \cite{6} and \cite{7} as a whole.\par
The most general action of $\mathcal N=1, D=4$ conformal supergravity coupled with SQCD described in gauge covariant conformal superspace formalism is
\begin{align}
S=&-3\int d^8z\,E\ \bar\Xi\,e^{-K/3}\,\Xi\nonumber\\
&\quad+\left(\int d^6z\,\E\ \Xi^3W(\Phi)-\frac14\int d^6z\,\E\ H_{(a)(b)}(\Phi)\W^{(a)\alpha}\W^{(b)}_\alpha+\hc\right).\label{action_conformal_SUGRA}
\end{align}
Here $d^8z\equiv d^4xd^4\theta, d^6z\equiv d^4xd^2\theta$, $\Xi$ is the chiral compensator, the only one chiral multiplet\footnote{In general we can consider a theory with two or more chiral multiplets with nonzero weights, but even for that case the theory is attributed to the action (\ref{action_conformal_SUGRA}) by field redifinition\cite{6}.} with nonzero conformal and chiral weights\footnote{The ratio between the conformal and chiral weights of a primary chiral superfield is fixed at 3/2.} $(\Delta, w)=(1,2/3)$, the real K\"ahler potential $K=K(\Phi, \bar\Phi)$, and the holomorphic superpotential $W(\Phi)$ and gauge kinetic coupling $H_{(ab)}(\Phi)$ are functions of the other chiral matter superfields $\Phi^i$ with zero weights $(\Delta, w)=(0,0)$, and the chiral gaugino multiplet $\W^{(a)}_\alpha$ is the gauge curvature multiplet of super-Yang-Mills. The chiral superfields $\Phi^I\equiv(\Xi, \Phi^i)$ satisfy the superconformal and gauge covariant chiral condition:
\begin{align}
\bar\nabla_{\dot\alpha}\Phi^I=0,
\end{align}
and primary condition:
\begin{align}
D\Phi^I=\Delta_I\Phi^I,\quad A\Phi=iw_I\Phi^I,\quad K_A\Phi^I=0,
\end{align}
where
\begin{align}
\nabla_M\equiv\del_M-\frac12\phi_M^{\phantom Mba}M_{ab}-B_MD-A_MA-f_M^{\phantom MA}K_A-\mathcal A_M^{(a)}X_{(a)}.
\end{align}
The operators $(M_{ab}, D, A, K_A)$ represent respectively Lorentz rotation, dilatation, chiral rotation, and special conformal boost in superconformal algebra, and the coefficients $(\phi_M^{\phantom Mab}, B_M, A_M, f_M^{\phantom MA})$ are the corresponding gauge fields. The last operator $X_{(a)}$ represents Yang-Mills gauge transformation, and the connection superfield $\mathcal A_M^{(a)}$ carries as many dynamical degrees of freedom as a real vector superfield $V$, after imposing the canonical curvature constraints
\begin{align}
\{\nabla_\alpha,\nabla_\beta\}=\{\bar\nabla_{\dot\alpha},\bar\nabla_{\dot\beta}\}=0,\quad\{\nabla_\alpha,\bar\nabla_{\dot\beta}\}=-2i\nabla_{\alpha\dot\beta}.
\end{align}
Solving the Bianchi identities under these constraints, the curvature operator $R_{AB}$ can be expressed in terms of a single superfield operator $\mathcal W_\alpha$ and its derivative:
\begin{align}
R_{\dot\alpha,\dot\beta\gamma}&\equiv-[\bar\nabla_{\dot\alpha},\nabla_{\dot\beta\gamma}]=2i\epsilon_{\dot\alpha\dot\beta}\W_\gamma,\\
R_{\alpha\dot\alpha,\beta\dot\beta}&\equiv-[\nabla_{\alpha\dot\alpha},\nabla_{\beta\dot\beta}]=-\epsilon_{\dot\alpha\dot\beta}\{\nabla_{(\alpha},\W_{\beta)}\}-\epsilon_{\alpha\beta}\{\bar\nabla_{(\dot\alpha},\bar\W_{\dot\beta)}\},
\end{align}
and $\W_\alpha$ is expressed in terms of the gravitational curvature multiplet $W_{\alpha\beta\gamma}$ and the Yang-Mills curvature multiplet $\W^{(a)}_\alpha$:
\begin{align}
\W_\alpha=(\epsilon\sigma^{bc})^{\beta\gamma}W_{\alpha\beta\gamma}M_{cb}+\frac12\nabla^\gamma W_{\gamma\alpha}^{\phantom{\gamma\alpha}\beta}S_\beta-\frac12\nabla^{\gamma\dot\beta}W_{\gamma\alpha}^{\phantom{\gamma\alpha}\beta}K_{\beta\dot\beta}+\W_\alpha^{(a)}X_{(a)},
\end{align}
with $K_A=(K_a,S_\alpha,\bar S^{\dot\alpha})$. 
The gauge curvature $\W^{(a)}_\alpha$ is related to the connection $\mathcal A^{(a)}_M$ and the vector superfield $V=V^{(a)}X_{(a)}$ as
\begin{align}
\mathcal F^{(a)}_{\dot\alpha,\gamma\dot\beta}&=2i\W^{(a)}_\gamma,\\
\W^{(a)}_\alpha X_{(a)}&=-\frac14\bar D^2e^{-V}D_\alpha e^V,
\end{align}
where
\begin{align}
\mathcal F^{(a)}_{MN}&\equiv\del_M\mathcal A_N^{(a)}-\del_N\mathcal A_M^{(a)}-\mathcal A^{(b)}_N\mathcal A_M^{(c)}f_{(c)(b)}^{\phantom{(b)(c)}(a)},
\end{align}
and $D_A\equiv\nabla_A+\mathcal A_A^{(a)}X_{(a)}$ is the covariant derivative in the theory without super-Yang-Mills.

\par
The action (\ref{action_conformal_SUGRA}) enjoys K\"ahler symmetry, i.e. a symmetry under compensator redifinition:
\begin{align}
\Xi\mapsto e^\Sigma\Xi,\quad K\mapsto K+3\Sigma+3\bar\Sigma,\quad W\mapsto e^{-3\Sigma}W,
\end{align}
where $\Sigma=\Sigma(\Phi)$ is a holomorphic function of $\Phi$'s. The action (\ref{action_conformal_SUGRA}) also enjoys the superconformal symmetry. Superconformal transformation acts on each field as
\begin{align}
E\mapsto e^{-(\Omega+\bar\Omega)}E,\quad \E\mapsto e^{-3\Omega}\E,\nonumber\\
\Xi\mapsto e^\Omega\,\Xi,\quad\W^{(a)}_\alpha\mapsto e^{(3/2)\Omega}\,\W^{(a)}_\alpha,\label{superconformal_transformation}
\end{align}
where $\Omega$ is a chiral superfield parameter. Note that $\W^{(a)}_\alpha$ has a nonzero conformal weight, even though the corresponding potential vector superfield $V$ is of weight zero. This is because the covariant derivative $\nabla_A$ depends on the supervielbein, which is transformed by superconformal transformation. 

\section{Superconformal Anomaly}
As is usual in gravitational theories, the superconformal transformation (\ref{superconformal_transformation}) can be anomalous in quantum theory, i.e. incompatible with local supersymmetry, namely general coordinate transformation in superspace. However, actually the {\it pure} conformal supergravity theory (without Yang-Mills and matter multiplets other than the compensator) is found to be anomaly-free after some calculation. The general, gauged and matter-coupled theory (\ref{action_conformal_SUGRA}) is anomalous, only against the Yang-Mills gauge symmetry and the K\"ahler symmetry. In this section we calculate the anomaly, i.e. the expectation value of the classical conserved quantity, in the way shown in Section \ref{section_Fujikawa_method}.\par
First, for the transformation (\ref{superconformal_transformation}), the conserved quantity is obtained by variating only the chiral compensator:
\begin{align}
Q=\frac\Delta{\Delta\Omega}S[E, e^\Omega\Xi, \Phi, V]\bigg|_{\Omega=0}=-3\cdot\frac{-1}4\bar\nabla^2\bar\Xi\,e^{-K/3}\,\Xi+3\,\Xi^2W(\Phi).
\end{align}
Then calculate the corresponding anomaly. For this case, the K\"ahler and gauge-covariant operator $\mathcal O$ should be taken
\begin{align}
\mathcal O^2\Xi&=\frac1{16}\bar\nabla^2e^{K/3}\nabla^2e^{-K/3}\Xi\\
&=\left(\Box-\frac12C^\alpha\nabla_\alpha+\cdots\right)\Xi,
\end{align}
with
\begin{align}
C^\alpha&\equiv\W^{(a)\alpha}X_{(a)}+W(K)^\alpha,\\
W(K)^\alpha&\equiv-\frac14\bar\nabla^2e^{K/3}\nabla^\alpha e^{-K/3}=\frac1{12}\bar\nabla^2\nabla^\alpha K.
\end{align}
Finally we obtain the anomalous Ward-Takahashi identity for the superconformal transformation:
\begin{align}
&\frac{-1}4\bar\nabla^2\left\langle\bar\Xi\,e^{-K/3}\,\Xi\right\rangle-\Big\langle\Xi^2W(\Phi)\Big\rangle\\
=&-\frac1{192\pi^2}C^\alpha C_\alpha\\
=&-\frac1{192\pi^2}\left(W(K)^\alpha W(K)_\alpha+W^{(a)\alpha}W^{(a)}_\alpha\right),
\end{align}
where we have used $\Tr X^{(a)}X^{(b)}=\delta^{(a)(b)}$.

\section{Summary}
In this paper we have calculated superconformal anomaly in $\mathcal N=1,D=4$ conformal supergravity:
\begin{align}
\frac{-1}4\bar\nabla^2\left\langle\bar\Xi\,e^{-K/3}\,\Xi\right\rangle-\Big\langle\Xi^2W(\Phi)\Big\rangle=-\frac1{192\pi^2}\Big(W(K)^\alpha W(K)_\alpha+\Tr W^\alpha W_\alpha\Big),
\end{align}
where the left side is the conserved current corresponding to superconformal transformation, while $W(K)^\alpha$ and $W^\alpha$ are the K\"ahler curvature multiplet and the usual gaugino multiplet of super-Yang-Mills, respectively. In order to calculate this result, we have also established an anomaly calculation method by path integral measure in superspace, which we have checked reproduces the traditional Konishi anomaly, the supersymmetric version of chiral anomaly in gauge theory.\par
This anomaly calculation technique is now limited in $\mathcal N=1$ superspace, but, naturally, there are also many intriguing topics concerning quantum anomaly or explicit calculation of path integral in $\mathcal N=2$ or more extended supersymmetric theories\cite{8}. For example, mere chiral anomaly in $\mathcal N=2$ SQCD\footnote{Nonlinear sigma model with $\mathcal N=2$ supersymmetry naturally enjoys hyperk\"ahler structure, in contrast to the $\mathcal N=1$ case endowed with K\"ahler structure. See \cite{11,12,13} for the detail.}\cite{8,9,10} or superconformal anomaly in $\mathcal N=2$ conformal supergravity\cite{14} are worth examining. As more advanced subjects, supersymmetric localisation\cite{15,16,17,18}, i.e. a technique of path integral calculation around instanton background, or conformal invariance in {\it quantum} $\mathcal N=4$ super-Yang-Mills theory are also attractive. The author have already begun to investigate some $\mathcal N=2$ cases. Further studies are expected for these topics.

\appendix
\section{Coefficient $1/64\pi^2$\label{appendix_coefficient}}
The coefficient $1/64\pi^2$ in (\ref{Konishi_anomaly}) is sometimes replaced by $1/32\pi^2$ or $1/16\pi^2$ in other papers \cite{3}. This is merely a conventional problem. There are two reasons responsible for the extra factor 2. First, in some paper they define the curvature multiplet as
\begin{align}
\tilde W_\alpha\equiv-\frac18\bar D^2e^{-2V}D_\alpha e^{2V},
\end{align}
accompanied by the interaction term $\bar\Phi\,e^{2V}\Phi$. Second, the component fields in chiral multiplets is sometimes defined \cite{8} as
\begin{align}
\Phi=A(x_L)+2\psi(x_L)\theta+F(x_L)\theta^2,
\end{align}
instead of $\Phi=A(x_L)+\sqrt2\psi(x_L)\theta+F(x_L)\theta^2$ in this paper. Each of these causes gives birth to an extra factor 2, and then the prefactor can be multiplied by 2 or 4.

\end{document}